\documentclass[aps,prb,twocolumn,showpacs,groupedaddress,preprintnumbers,amsmath,amssymb]{revtex4}
\usepackage {graphicx}
\usepackage{color}

\begin {document}

\title {Pressure-driven metal-insulator transition in BiFeO$_3$ from Dynamical Mean-Field Theory}

\author {A.~O.~Shorikov}
\affiliation{M.N. Miheev Institute of Metal Physics of Ural Branch of Russian Academy of Sciences - 620137 Yekaterinburg, Russia}
\affiliation{Ural Federal University - 620002 Yekaterinburg, Russia}

\author {A.~V.~Lukoyanov}
\affiliation{M.N. Miheev Institute of Metal Physics of Ural Branch of Russian Academy of Sciences - 620137 Yekaterinburg, Russia}
\affiliation{Ural Federal University - 620002 Yekaterinburg, Russia}

\author {V.~I.~Anisimov}
\affiliation{M.N. Miheev Institute of Metal Physics of Ural Branch of Russian Academy of Sciences - 620137 Yekaterinburg, Russia}
\affiliation{Ural Federal University - 620002 Yekaterinburg, Russia}

\author {S.~Y.~Savrasov}
\affiliation {Department of Physics, University of California, Davis, California 95616, USA}

\begin {abstract}

A metal--insulator transition (MIT) in BiFeO$_3$ under pressure was investigated by a method combining Generalized Gradient Corrected Local Density Approximation with Dynamical Mean--Field Theory (GGA+DMFT). Our paramagnetic calculations are found to be in agreement with experimental phase diagram: Magnetic and spectral properties of BiFeO3 at ambient and high pressures were calculated for three experimental crystal structures $R3c$, $Pbnm$ and $Pm\bar{3}m$. At ambient pressure in the $R3c$ phase, an insulating gap of 1.2 eV was obtained in good agreement with its experimental value. Both $R3c$ and $Pbnm$ phases have a metal--insulator transition that occurs simultaneously with a high--spin (HS) to low--spin (LS) transition. The critical pressure for the $Pbnm$ phase is 25--33 GPa that agrees well with the experimental observations. The high pressure and temperature $Pm\bar{3}m$ phase exhibits a metallic behavior observed experimentally as well as in our calculations in the whole range of considered pressures and undergoes to the LS state at 33 GPa where a $Pbnm$ to $Pm\bar{3}m$ transition is experimentally observed. The antiferromagnetic GGA+DMFT calculations carried out for the $Pbnm$ structure result in simultaneous MIT and HS-LS transitions at a critical pressure of 43 GPa in agreement with the experimental data.

\end {abstract}

% 71.30.+h	Metal-insulator transitions and other electronic transitions
% 71.27.+a Strongly correlated electron systems; heavy fermions
% 71.20.-b	Electron density of states and band structure of crystalline solids

\maketitle
\section{Introduction}
Multiferroics are used in various applications for energy production, transmission of high voltage lines, data storage devices, and sensors~\cite{tokura06}. They will help to replace a number of currently used lead-based materials which contain lead toxic and harmful to the environment. One of the most promising candidates for applications, bismuth ferrite BiFeO$_3$, is actively studied because of the coupling between ferroelectric and magnetic order around room temperature in this compound. Recent investigations of BiFeO$_3$ at high pressures up to 60 GPa reveal metal-insulator~\cite{gavriluk2008_2} and high-spin to low-spin (HS-LS) transitions~\cite{gavriluk2008} in Fe$^{3+}$ at room temperature in a relatively wide pressure range 40-55 GPa accomplished by the structural phase transition~\cite{gavriluk2007}.

The problem becomes even more complicated taking into account that BiFeO$_3$ has a rich phase diagram. At ambient pressure and up to 1100~K it has rhombohedral ($R3c$) crystal structure\cite{moreau1971}. Increasing temperature and pressure, the structure of BiFeO$_3$ transforms into orthorhombic $Pbnm$~\cite{Arnold2009} and then cubic $Pm\bar{3}m$~\cite{Redfern2009}. 

For many years, metal-insulator transition (MIT) in $d$ or $f$ metal compounds~\cite{Imada} is one of the central issues in condensed matter physics. The most spectacular examples are pressure-driven transitions from a wide gap Mott insulator to metallic state in transition metal oxides. For MnO and Fe$_2$O$_3$ ($d^5$ configuration), the metal-insulator transition is accompanied by the high-spin to low-spin (HS-LS) transition. 

Recently, these and other MITs and spin transitions were successfully described theoretically \cite{mno,fe2o3,feo,fesi} employing the method combining density functional approximations (like, GGA or LDA) with dynamical mean-field theory~\cite{GGA+DMFT}. Pressure-driven MIT correlated with magnetic collapse could be treated as a delocalization of magnetic electrons or structural phase transition into new phase with N\'eel point below room temperature~\cite{gavriluk2005}. 

Another Mott-type mechanism controlled by dramatic $U_{eff}$ decrease due to HS-LS transition was proposed in Ref.\cite{gavriluk2008_2}. Mott-type MIT driven by a broadening of the $t_{2g}$ states was confirmed by the LDA+$U$ calculations which revealed the HS-LS transition at 36 GPa and a transition to the metallic phase with no localized moment above 72 GPa~\cite{Gonzalez2009}.

In this paper we investigate the properties of BiFeO$_3$ under pressure employing the method accounting for dynamical electronic correlations and considering three experimental crystal structures.

\section{Method}
The GGA+DMFT method \cite{GGA+DMFT} is realized in a computational scheme constructed in the following way: first, a Hamiltonian $\hat H_{GGA}$ is produced using converged GGA results for a compound under investigation, then the many-body Hamiltonian is set up, and finally the corresponding self-consistent DMFT equations are solved. In this work the Hamiltonians $\hat H_{GGA}$ are constructed in a Wannier function (WF) basis~\cite{Wannier37, MarzariVanderbilt} using the projection procedure described in detail in Ref.~\onlinecite{Korotin}. Initial ab-initio calculations of the electronic structure are done within the pseudo-potential plane-wave method, as implemented in Quantum ESPRESSO~\cite{PW}. 

The WFs are defined by the choice of Bloch functions Hilbert space and by a set of trial localized orbitals that will be projected on these Bloch functions. The basis set includes all bands  that are formed by O 2$p$  and Fe 3$d$ states and correspondingly the full set of the O 2$p$ and Fe 3$d$ atomic orbitals to be projected on Bloch functions for these bands. That would correspond to the extended model where in addition to the $d$ orbitals all $p$ orbitals are also included. 

The resulting $p-d$ Hamiltonian to be solved by DMFT has the form
\begin{equation}
\hat H= \hat H_{GGA}- \hat H_{dc}+\frac{1}{2}\sum_{i,\alpha,\beta,\sigma,\sigma^{\prime}}
U^{\sigma\sigma^{\prime}}_{\alpha\beta}\hat n^{d}_{i\alpha\sigma}\hat n^{d}_{i\beta\sigma^{\prime}},
\label{eq:ham}
\end{equation}
where $U^{\sigma\sigma^{\prime}}_{\alpha\beta}$ is the Coulomb interaction matrix, $\hat n^d_{i\alpha\sigma}$ is the occupation number operator for the $d$ electrons with orbitals $\alpha$ or $\beta$ and spin indices $\sigma$ or $\sigma^{\prime}$ on the $i$-th site. The term $\hat H_{dc}$ stands for the $d$-$d$ interaction already accounted for in GGA, so called double-counting correction. 
In the present calculation the double-counting was chosen in the following form $\hat H_{dc}=\bar{U}(n_{\rm dmft}-\frac{1}{2})\hat{I}$. Here $n_{\rm dmft}$ is the self-consistent total number of $d$ electrons obtained within GGA+DMFT, $\bar{U}$ is the average Coulomb parameter for the $d$ states, and $\hat I$ is unit operator. 

The elements of $U_{\alpha\beta}^{\sigma\sigma'}$ matrix are parameterized by $U$ and $J_H$ according to the procedure described in \cite{LichtAnisZaanen}. The values of Coulomb repulsion parameter $U$ and Hund exchange parameter $J_H$ were calculated by the constrained LDA method \cite{U-calc} on Wannier functions \cite{Korotin}. The values $U$=6 eV and $J_H$=0.93 eV obtained in these calculations are close to the previous estimations~\cite{Savrasov, Gonzalez2009}. The effective impurity problem in DMFT was solved by the hybridization expansion Continuous-Time Quantum Monte-Carlo method (CT-QMC) \cite{CTQMC}. The calculations for all volumes were performed in the paramagnetic state for inverse temperature $\beta=1/T$=15 eV$^{-1}$ corresponding to 770~K. According to the phase diagram, no long-range ordering in BiFeO$_3$ was found in this temperature region. The spectral functions on real energies were calculated employing Maximum Entropy Method (MEM)\cite{mem}.

\section{Results and discussion}
In all structures under investigation iron is surrounded by oxygen ions forming an octahedron. In the high pressure cubic $Pm\bar{3}m$  phase, the Fe $d$ band is split by crystal field into threefold degenerate $t_{2g}$ and twofold degenerate $e_g$ sub-bands. A trigonal distortion in the $R3c$ structure lowers the point symmetry group $O_h$ to $D_{3h}$, so that threefold degenerate $t_{2g}$ becomes split into twofold degenerate $e_g^\pi$ and non-degenerate $a_{1g}$ band. The $Pbnm$ structure has the lowest symmetry and all orbitals become nonequivalent but still three groups of orbitals could be considered: $e_{g1}^\sigma$ and $e_{g2}^\sigma$,  $e_{g1}^\pi$ and  $e_{g2}^\pi$ and $a_{1g}$. For simplicity we will use cubic $t_{2g}$ and $e_g$ orbitals notations for analysis hereafter. 

Previous calculations demonstrated that LDA fails to describe an insulating character of the BiFeO$_3$ ground state at ambient pressure and for all volumes it is metallic in all structures. However, a gap of the AFM origin was obtained in the GGA calculation made for several magnetic structures~\cite{Ravindran2006}.

\begin {figure}
\vspace{5mm}
\includegraphics [width=0.425\textwidth,clip=true]{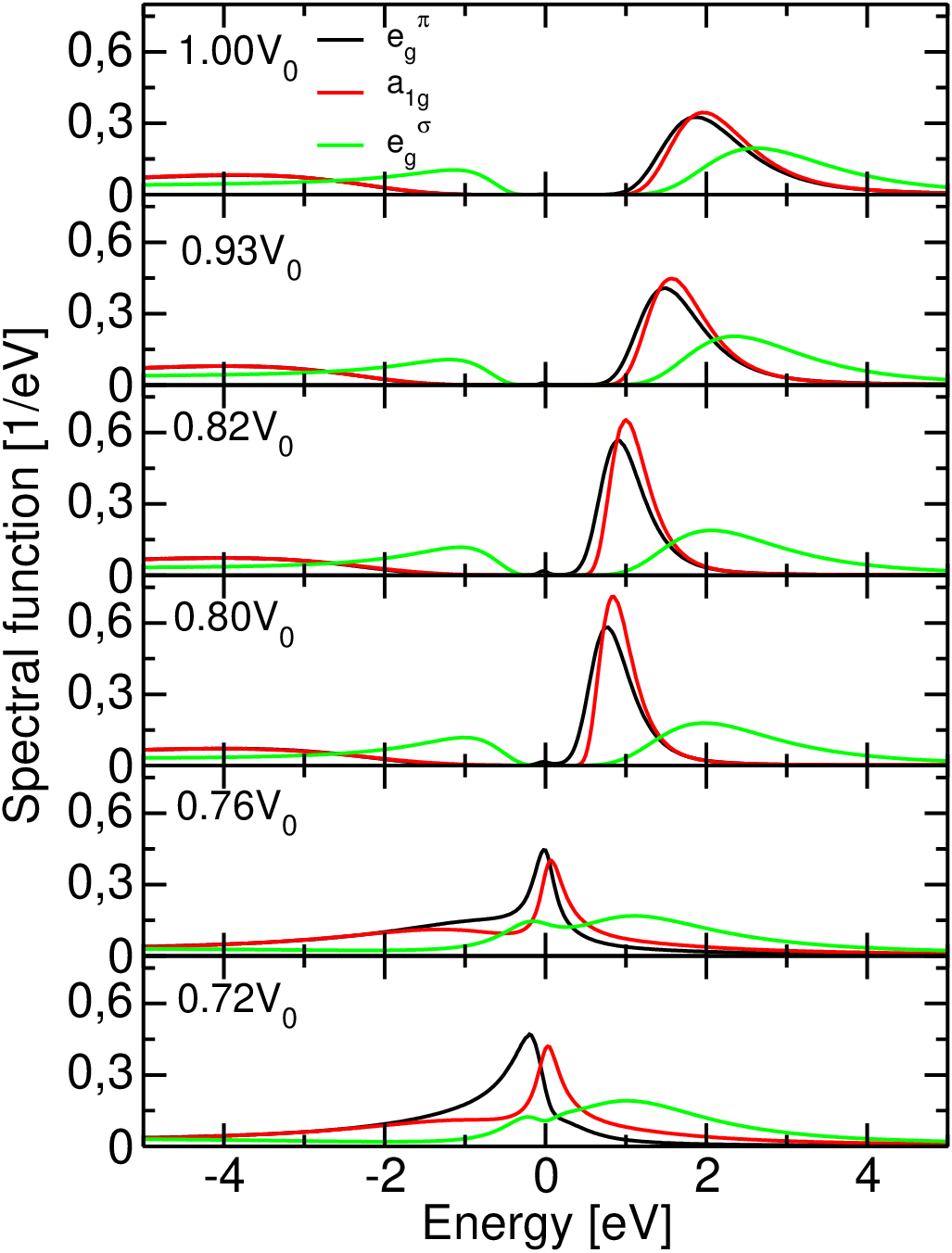}
\caption {(Color online) Spectral function of Fe $d$ states for different volumes related to the ambient pressure volume $V_0$ obtained in the GGA+DMFT (CT-QMC) calculations for the low pressure $R3c$ phase at 770 K.}
\label {fig:R3c_sf}
\end {figure}

Our GGA+DMFT calculations produced for the $R3c$  phase of BiFeO$_3$ show that taking into account Coulomb correlation effects results a wide-gap Mott insulator and high-spin state for ambient pressure in agreement with the experimental data. The obtained spectral functions for different cell volumes are shown in Fig.~\ref{fig:R3c_sf}. The calculated energy gap value of about 1.2 eV agrees well with the  experimental value of 2.4 eV for direct optical gap~\cite{Ramachandran2012,Pisarev2009} and 1.8 eV indirect optical gap at AP~\cite{Fruth2007} at room temperature and indirect gap value 1.3 eV measured in thin films~\cite{Gujar2007}, 1.5 eV at 820K~\cite{Palai2008}.  
The magnitude of magnetic moment $\sqrt{<m_z^2>}$ is 4.6$\mu_B$ at AP. This number agrees very well with the high-spin state of the Fe$^{+2}$ ion (d$^5$ configuration) in cubic crystal field: 2 electrons in the $e_{g}$ band and 3 electrons in the $t_{2g}$ band with the magnetic moment value of 5 $\mu_B$. Then the magnetic moment decreases and at 0.76$V_0$ drops down to 2.5 $\mu_B$, and it equals to 2.0 $\mu_B$ when the volume is 0.72$V_0$ (63 GPa). The later value is close to the one expected for the LS state with 5 electrons in the $t_{2g}$ band but strong hybridization between iron and oxygen leads to the non zero occupation of the $e_g$ orbitals. This mechanism was obtained in all three crystal structures and is considered in detail for the $Pbnm$ case.

\begin {figure}
\vspace{5mm}
\includegraphics [width=0.425\textwidth,clip=true]{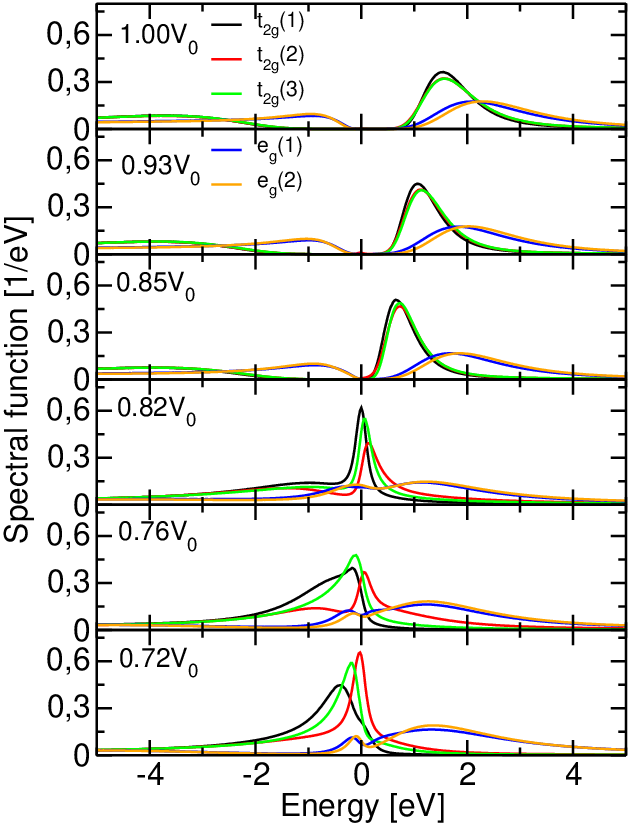}
\caption {(Color online) Spectral function of Fe-d states vs. volume reduction obtained in GGA+DMFT (CT-QMC) calculations for the $Pbnm$ phase at 770K.}
\label {fig:Pbnm_sf}
\end {figure}
 
\begin {figure}
\vspace{5mm}
\includegraphics [width=0.425\textwidth,clip=true]{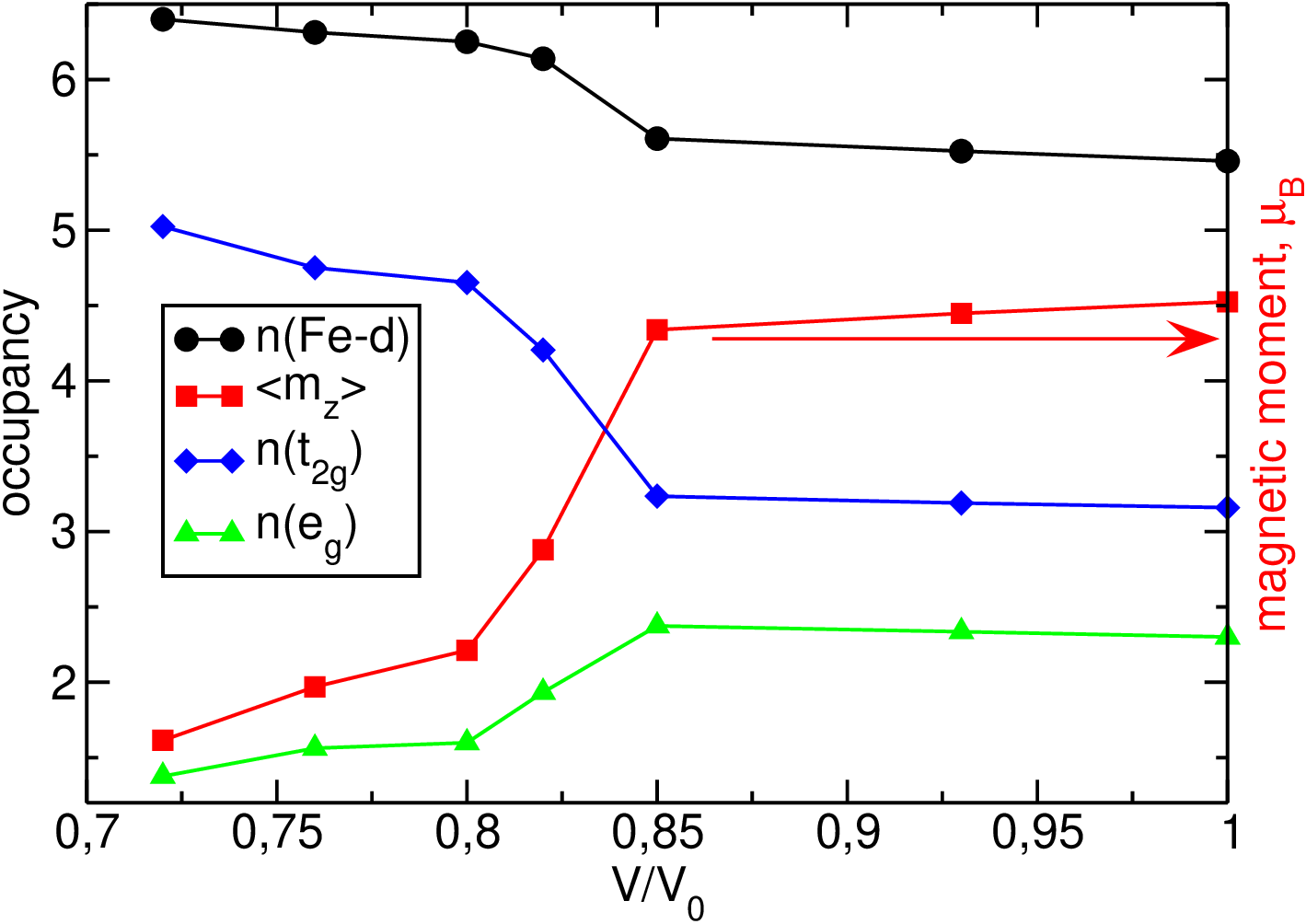}
\caption {(Color online) Magnetic moments (red square), occupancy of the $d$ states (black circles), $t_{2g}$ (blue diamonds), and $e_g$ (green triangles) states obtained in the GGA+DMFT (CT-QMC) calculations for the $Pbnm$ phase. }
\label {fig:mom}
\end {figure}

The most interesting phase is $Pbnm$ since both the HS-LS and metal-insulator transitions occur simultaneously with the $Pbnm$ to $Pm\bar{3}m$ phase transition. The GGA+DMFT spectral functions for different cell volumes are shown in Fig.~\ref{fig:Pbnm_sf}. Note, that the $Pbnm$ phase doesn't exist to ~2 GPa at 770 K. But BiFeO$_3$ in the $Pbnm$ structure and experimental with cell volume of $R3c$ at ambient pressure is an insulator with a gap of $~$1 eV which is close to the experimental data~\cite{Pisarev2009, Fruth2007, Palai2008}. When the cell volume is 0.85V$_0$ the gap is closed but iron is still in the HS state (see Fig.~\ref{fig:mom}) that contradicts to the mechanism proposed in Ref.~\onlinecite{gavriluk2008_2}. Then together with the volume decrease the $t_{2g}$ occupancy increases and both magnetic moment and a number of $e_{g}$ electrons shrink. Both transitions don't appear to happen instantly but have a crossover region 0.85-0.82V$_0$ that agrees with the experimental observations~\cite{catalan2009, gavriluk2008_2, gavriluk2008}.   

MIT at room temperature where long range order exists occurs in the pressure range 45-55 GPa. In order to reproduce the temperature dependence of critical pressure value the paramagnetic GGA+DMFT calculation for $\beta$=40eV$^{-1}$ eV was carried out. The results obtained show that at 33 GPa BiFeO$_3$ is already in the metallic LS state as well as at higher temperature. Hence increase of the critical pressure isn't a temperature-driven effect. Then we have carried out magnetic GGA+DMFT calculation for the same cell volumes of $Pbnm$ structures.  Small external magnetic field 0.01 eV was applied to each Fe atom so that direction of the field on sites corresponds to experimentally observed G-type AFM. 
The gap value increases slightly but more important magnetic solution was stabilized and MIT occurs between 0.82-0.8 V$_0$ that corresponds to 33-43 GPa. This value is close to transition pressure 45-55 GPa measured at 300 K in AFM phase. 

\begin {figure}
\vspace{5mm}
\includegraphics [width=0.425\textwidth,clip=true]{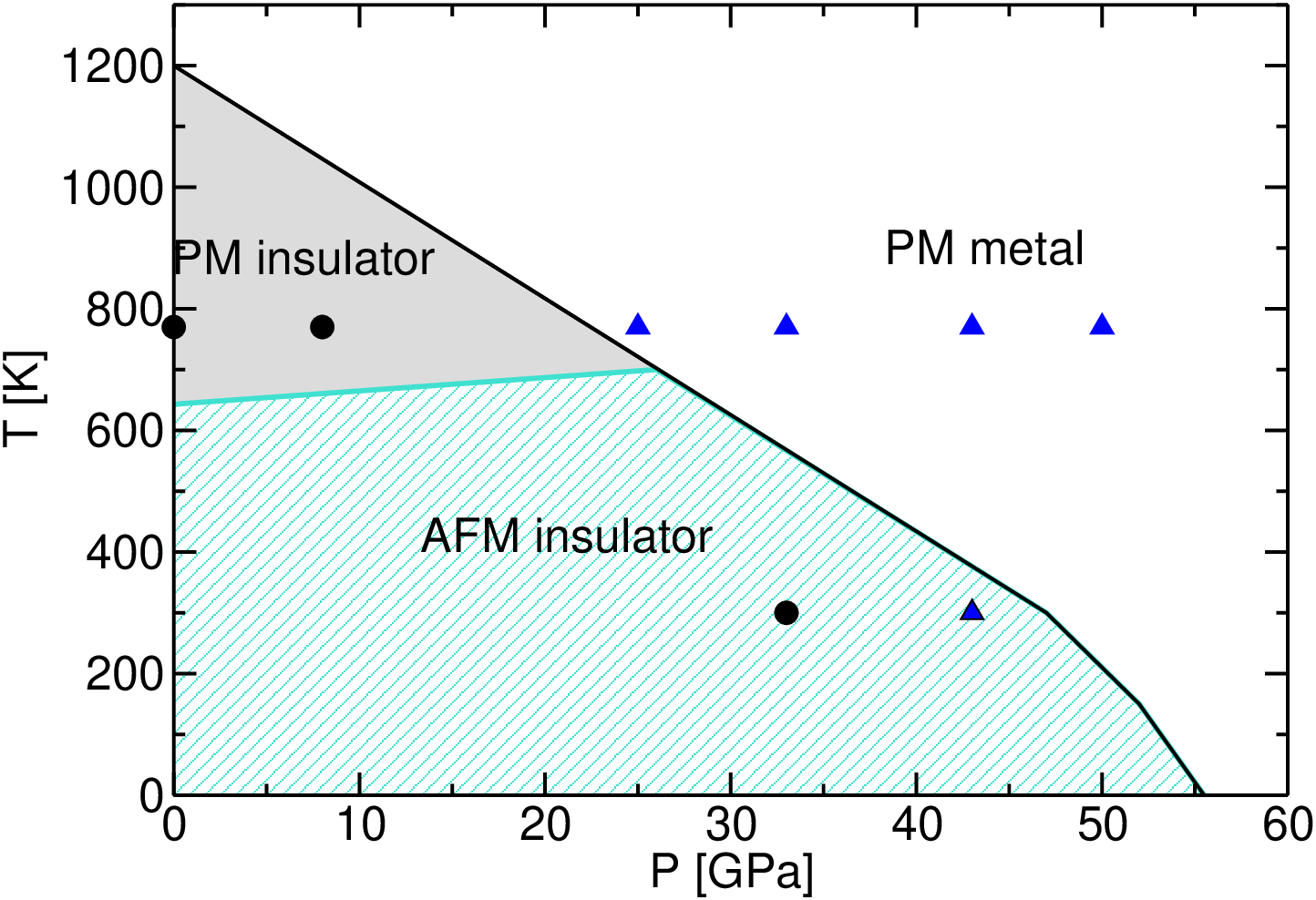}
\caption {(Color online) Combined phase diagram of BiFeO$_3$. Experimental data taken from from Ref.\onlinecite{catalan2009} are shown with solid lines. Insulator and metallic solutions obtained in GGA+DMFT calculation are shown with black squares and blue triangles, correspondingly. Critical pressure obtained in paramagnetic calculation is 25 GPa. Spin polarized GGA+DMFT results at 300 K results in IMT in a range 33-43 GPa.}
\label {fig:phase_diag}
\end {figure}

\begin {figure}
%\vspace{5mm}
\includegraphics [width=0.425\textwidth,clip=true]{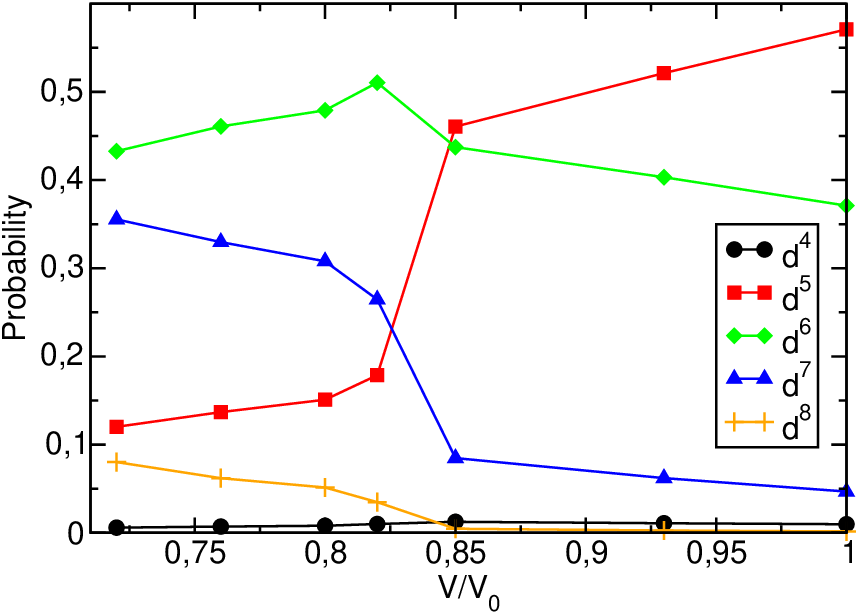}
%\vspace{25mm}
\includegraphics [width=0.425\textwidth,clip=true]{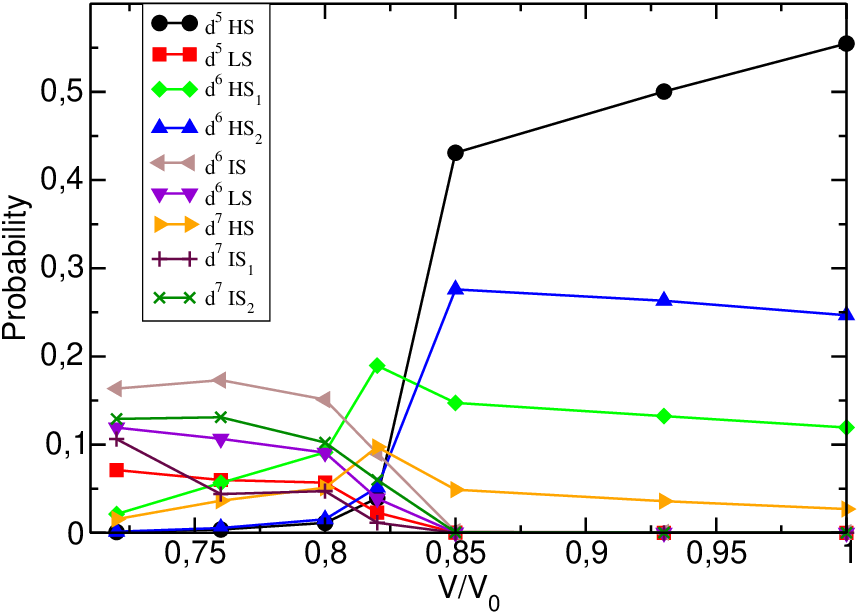}
\caption {(Color online)Probabilities of different charge states $d^n$  (upper panel) and orbital configurations of Fe ion (lower panel) vs. pressure obtained in the GGA+DMFT (CT-QMC) calculations for $Pbnm$ phase (shown only largest ones): 
$d^{5}$-HS     (up: $t_{2g}^{3}e_{g}^{2}$, down: $t_{2g}^{0}e_{g}^{0}$),
$d^{5}$-LS     (up: $t_{2g}^{3}e_{g}^{0}$, down: $t_{2g}^{2}e_{g}^{0}$),
$d^{6}$-HS$_1$ (up: $t_{2g}^{3}e_{g}^{2}$, down: $t_{2g}^{1}e_{g}^{0}$), 
$d^{6}$-HS$_2$ (up: $t_{2g}^{3}e_{g}^{2}$, down: $t_{2g}^{0}e_{g}^{1}$),
$d^{6}$-IS     (up: $t_{2g}^{3}e_{g}^{1}$, down: $t_{2g}^{2}e_{g}^{0}$), 
$d^{6}$-LS     (up: $t_{2g}^{3}e_{g}^{0}$, down: $t_{2g}^{2}e_{g}^{1}$),
$d^{7}$-HS     (up: $t_{2g}^{3}e_{g}^{2}$, down: $t_{2g}^{1}e_{g}^{1}$),
$d^{7}$-IS$_1$ (up: $t_{2g}^{3}e_{g}^{1}$, down: $t_{2g}^{3}e_{g}^{0}$),
$d^{7}$-IS$_2$ (up: $t_{2g}^{3}e_{g}^{1}$, down: $t_{2g}^{2}e_{g}^{1}$).}
\label {fig:prob}
\end {figure}

Probabilities of charge and orbital configurations measured in the GGA+DMFT method within impurity solver are shown in Fig.\ref{fig:prob}. The hybridization between iron and oxygen and hence the $d$ states occupation increases with the volume contraction. At ambient pressure, the $d^5$ configuration has the larger impact but the $d^6$ probability is also sizable. With the volume reduction probabilities of $d^6$ and $d^7$ configurations grow up and the probability of the $d^5$ configuration shrinks, see the upper panel of Fig.\ref{fig:prob}. 

More detailed picture of orbital configuration impacts is shown in Fig.~\ref{fig:prob} (lower panel). One can see that the probabilities of the HS configuration with any number of electrons ($d^5$, $d^6$ and $d^7$) which dominate at ambient pressure goes down and after the cell volume become smaller than 0.85V$_0$ drops and LS and IS configuration probability increase.   

The calculation carried out for the high temperature cubic $Pm\bar{3}m$ phase demonstrates that for all  experimental volumes BiFeO$_3$ has a metallic behavior. The high spin to low spin state transition was also obtained in the calculation for the $Pm\bar{3}m$ phase at 0.8V$_0$ (33 GPa). 

\section{In conclusion}
We have performed the GGA+DMFT calculations for BiFeO$_3$ at 700 K where it has no long range order for a number of different cell volumes corresponding to the whole range of experimentally applied pressures (from ambient pressure up to 60 GPa).  In agreement with experiments, the spectral functions for BiFeO$_3$ in the $R3c$ phase as well as in $Pbnm$ at ambient pressure demonstrates an energy gap of about 1.2 eV in a good agreement with experiment. When the unit cell volume is  0.85-0.82 V$_0$ that corresponds to 25-33 GPa BiFeO$_3$ in the $Pbnm$ crystal structure becomes metallic, and, simultaneously, its magnetic moments shrink. In the low pressure $R3c$ structure, the same transition occurs in cell with the smaller volume 0.8-0.76 V$_0$ (40-50 GPa). The high pressure cubic $Pm\bar{3}m$ phase is metallic in the whole range of the investigated pressures (cell volumes) but the HS-LS transition also takes place when the cell volume is smaller than 0.82 V$_0$ (33 GPa). The higher critical pressure of 43 GPa was obtained in the AFM GGA+DMFT calculations for the $Pbnm$ structure in good agreement with the experimental phase diagram.

\section{Acknowledgments} 
The authors thank P.~Werner for the CT-QMC impurity solver. The GGA+DMFT calculations were performed on the Supercomputing center of IMM UrB RAS. This work was partially supported by RFBR (project No. 13-02-00050). This publication is based on work supported by a grant from the U.S. Civilian Research \& Development Foundation (CRDF Global) via a joint grant No. RUP1-7077-EK-12 (12-CD-2).


\begin{thebibliography}{99}
%Multiferroics
\bibitem{tokura06} Y. Tokura, Science {\bf312}, 1481 (2006).

%Mott-type MIT
\bibitem{gavriluk2008_2} A.G. Gavriliuk {\it et al.},
% V. V. Struzhkin, I. S. Lyubutin, S. G. Ovchinnikov, M. Y. Hu, P. Chow, 
Phys. Rev. B {\bf77}, 155112 (2008).

\bibitem{gavriluk2008} I.S. Lyubutin, A.G. Gavriluk, V.V. Struzhkin, JETP Lett {\bf 88} 524 (2008).

\bibitem{gavriluk2007} A.G. Gavriluk {\it et al.}, JETP Lett {\bf 86} 197 (2007).

%MIT
\bibitem {Imada} M. Imada, A. Fujimori, and Y. Tokura, Rev. Mod. Phys. \textbf {70}, 1039 (1998).

\bibitem {GGA+DMFT} V.I. Anisimov {\it et al.}, 
%, A. I. Poteryaev, M. A. Korotin, A. O.
%Anokhin, and G. Kotliar
J. Phys.: Condens. Matter \textbf {9}, 7359
(1997); A. I. Lichtenstein and M. I. Katsnelson, Phys. Rev. B \textbf {57}, 6884 (1998); K. Held
%, I. A. Nekrasov, G. Keller, V. Eyert, N. Bl\"umer, A.
%K. McMahan, R. T. Scalettar, Th. Pruschke, V. I. Anisimov, and D.
%Vollhardt
{\it et al.}, Phys. Stat. Sol. (b) \textbf {243}, 2599 (2006).

\bibitem{mno} J. Kunes {\it et al.},
% Alexey V. Lukoyanov, Vladimir I. Anisimov, Richard T. Scalettar, Warren E. Pickett, 
Nature Materials \textbf {7}, 198 (2008).

\bibitem{fe2o3}	J. Kunes {\it et al.}, 
%Dm. M. Korotin, M. A. Korotin, V. I. Anisimov, and P. Werner, 
Phys. Rev. Lett. \textbf {102}, 146402 (2009).
\bibitem{feo}	A.O. Shorikov {\it et al.}, 
%Z.V. Pchelkina,  V. I. Anisimov, S.L. Skornyakov, M. A. Korotin
Phys. Rev. B \textbf {82}, 195101 (2010).

\bibitem{fesi} V. V. Mazurenko {\it et al.}, 
%A. O. Shorikov, A. V. Lukoyanov, K. Kharlov, E. Gorelov, A. I. Lichtenstein, and V. I. Anisimov,
Phys. Rev. B \textbf {81}, 125131 (2010).

\bibitem{moreau1971} M. Moreau {\it et al}., J. Phys. Chem. Solids {\bf 32}, 1315 (1971).

\bibitem{Arnold2009} D. C. Arnold {\it et al},
% K. S. Knight, F. D. Morrison, P. Lightfoot, 
Phys. Rev. Lett.  {\bf 102}, 027602 (2009).

\bibitem{Redfern2009} S. A. T. Redfern {\it et al},
% J. N. Walsh, S. M. Clark, G. Catalan, J. F. Scott, 
arXiv:0901.3748, 2009.

\bibitem{catalan2009} G. Catalan, J. Scott, Advanced Materials. {\bf 21}, 2463-2485 (2009).

\bibitem{gavriluk2005} A. G. Gavriliuk {\it et al}, 
%V. V. Struzhkin, I. S. Lyubutin, M. Y. Hu, H. K.Mao, 
JETP Lett.{\bf 82}, 224 (2005).

%ab initio
\bibitem{Gonzalez2009} O. E. Gonzalez-Vazquez, J. In˜iguez, Phys Rev. B  {\bf79}, 064102 (2009).

\bibitem {PW} S. Baroni, S. de Gironcoli, A. D. Corso, and P. Giannozzi,
http://www.pwscf.org.

\bibitem {Wannier37} G. H. Wannier, Phys. Rev. \textbf {52}, 191 (1937).

\bibitem {MarzariVanderbilt} N. Marzari and D. Vanderbilt, Phys. Rev. B
\textbf {56}, 12847 (1997); W. Ku, H. Rosner, W. E. Pickett, and R. T.
Scalettar, Phys. Rev. Lett. \textbf {89}, 167204 (2002).

\bibitem{Korotin} Dm. Korotin {\it et al}.,
Euro. Phys. J. B {\bf 65}, 1434 (2008).

\bibitem{LichtAnisZaanen} A. I. Liechtenstein, V. I. Anisimov, and J. Zaanen,
Phys. Rev. B {\bf 52}, R5467 (1995).

\bibitem {U-calc} P. H. Dederichs{\it et al}.,
% S. Bl\"ugel, R. Zeller, and H. Akai,
Phys. Rev. Lett. \textbf {53}, 2512 (1984); O. Gunnarsson{\it et al}.
%, O. K. Andersen, O. Jepsen, and J. Zaanen, 
Phys. Rev. B \textbf {39}, 1708 (1989); V. I. Anisimov and O. Gunnarsson, \textit {ibid.} \textbf {43}, 7570 (1991).

% U for FeO 
\bibitem {Savrasov} S.A. Gramsch, R.E. Cohen and S.Yu. Savrasov, Amer. Miner. \textbf {88}, 257-261 (2003). 

\bibitem {CTQMC} P. Werner {\it et al}., 
Phys. Rev. Lett. {\bf 97}, 076405 (2006).

\bibitem{mem} Mark Jarrell and J. E. Gubernatis, Phys. Rep. {\bf 269}, 133 (1996).

\bibitem{Ravindran2006} P. Ravindran {\it et al}
%, R. Vidya, A. Kjekshus, H. Fjellva  O. Eriksson 
Phys. Rev. B {\bf 74}, 224412 (2006).

\bibitem{Ramachandran2012} B. Ramachandrana, and M. S. Ramachandra Rao, J. Appl. Phys. {\bf 112}, 073516 (2012).

\bibitem{Pisarev2009} R. V. Pisarev, A. S. Moskvin, A. M. Kalashnikova, and Th. Rasing, Phys. Rev. B {\bf79}, 235128 (2009).

%Thin films 
\bibitem{Fruth2007} V. Fruth {\it et al}
%, E. Tenea, M. Gartner, M. Anastasescu, D. Berger, R. Ramer, and M. Zaharescu, 
J. Eur. Ceram. Soc. {\bf27}, 937 (2007).

\bibitem{Palai2008} R. Palai {\it et al},
% R. S. Katiyar,H. Schmid, P. Tissot, S. J. Clark, J. Robertson, S. A. T. Redfern, G. Catalan, J. F. Scott, 
Phys. Rev. B  {\bf 77}, 014110 (2008).
%Thin films 
\bibitem{Gujar2007} T. P. Gujar, V. R. Shinde, C. D. Lokhande, Mater. Chem. Phys.  {\bf103}, 142 (2007). 

\end{thebibliography}
\end {document}